\documentclass{article}
\usepackage{spconf,amsmath,graphicx}
\usepackage{tabularx,booktabs}
\newcolumntype{C}{>{\centering\arraybackslash}X} 

\graphicspath{{./images/}}

\title{Fine-grained MRI Reconstruction using Attentive Selection Generative Adversarial Networks}
\name{Jingshuai Liu \qquad Mehrdad Yaghoobi}
\address{
    IDCOM, School of Engineering, University of Edinburgh, UK, EH9 3JE\\
    email: s1678560@ed.ac.uk,  m.yaghoobi-vaighan@ed.ac.uk
}

\begin{document}

“© 2021 IEEE.  Personal use of this material is permitted.  Permission from IEEE must be obtained for all other uses, in any current or future media, including reprinting/republishing this material for advertising or promotional purposes, creating new collective works, for resale or redistribution to servers or lists, or reuse of any copyrighted component of this work in other works.”
\newpage

\maketitle
\begin{abstract}
    Compressed sensing (CS) leverages the sparsity prior to provide the foundation for fast magnetic resonance imaging (fastMRI). However, iterative solvers for ill-posed problems hinder their adaption to time-critical applications. Moreover, such a prior can be neither rich to capture complicated anatomical structures nor applicable to meet the demand of high-fidelity reconstructions in modern MRI. 
    
    Inspired by the state-of-the-art methods in image generation, we propose a novel attention-based deep learning framework to provide high-quality MRI reconstruction. We incorporate large-field contextual feature integration and attention selection in a generative adversarial network (GAN) framework. We demonstrate that the proposed model can produce superior results compared to other deep learning-based methods in terms of image quality, and relevance to the MRI reconstruction in an extremely low sampling rate diet.
\end{abstract}
\section{Introduction}
Magnetic resonance imaging (MRI) utilizes a strong magnetic field and radio waves to primarily generate 2-dimensional (2-D) slices of cross-sections, and provides a radiation-free diagnosis tool. However, the time-consuming sampling and reconstruction steps impede its applications in time-critical diagnosis. The long acquisition time can be reduced via parallel imaging (PI). One successful PI method is to estimate the unobserved data in $k$-space from their neighboring points, by using a GRAPPA kernel \cite{grappa} which is estimated using the fully collected data in the central region. However, such methods require expensive equipments and it is difficult to remove strong aliasing artifacts, using traditional PI methods.

Model-based MRI reconstruction leverages compressive sampling methods based on the data sparsity assumption. They model the image prior in the form of signal sparsity in some domain, e.g. Fourier space, and achieve accurate reconstruction by solving nonlinear optimizations, which are successful if some conditions, e.g. restricted isometry property, are met. The challenge in meeting such hypotheses in real-world scenarios hinders the development of CS methods in fast MRI reconstructions. In a contrast to the request for sparsity of signals in image domain, or a transform domain \cite{blind_cs_mri}, \cite{dl_mri} proposes to learn a sparse basis via dictionary learning which enables more parsimonious representations. However, the limited capacity of the sparsity prior puts restrictions on fast MR imaging and leaves rooms for improvements using deep learning-based methods.

Deep neural networks show great effectiveness in feature representation and have been leveraged to recover the under-sampled MRI observations. One parallel MR imaging method is proposed in \cite{grappa_net} using two U-shaped networks (U-net) to predict the missing pixels. Generative adversarial networks (GAN) \cite{vallina_gan} show success in image generation and can be potentially employed in MRI reconstructions. The method in \cite{inverse_gan_mri} performs MRI reconstruction by jointly optimizing the latent space of a pre-trained GAN to enforce it being in agreement with the measurements and tuning the parameters of the model to leverage the deep image prior \cite{dip}. Motivated by the previous works in domain translation \cite{cyclegan}, concurrent researches using cyclic adversarial frameworks include: RefineGAN in \cite{mri_reconstruction_with_cyclic_losses} which applies the cyclic consistency to learn the mapping of the under-sampled measurements, and \cite{cyclegan_inverse} which utilizes the theory of optimal transport to provide a foundation for cycle-consistent GAN (CycleGAN).


We propose to learn the mapping from the under-sampled data to the alias-free images by leveraging the generative prior and cyclic data consistency. We introduce a novel deep de-aliasing module to capture large-field spatial dependencies in feature spaces and form a coarse to fine MR image mapping in two stages. Our work is the first endeavour to present a channel-wise and spatial attention selection for MRI reconstruction. By qualitative and quantitative evaluations, we demonstrate that the proposed framework outperforms other data-driven methods in terms of reconstruction quality.

\begin{figure*}[h]

    \begin{minipage}[b]{1.0\linewidth}
        \centering
        \centerline{\includegraphics[width=16.cm]{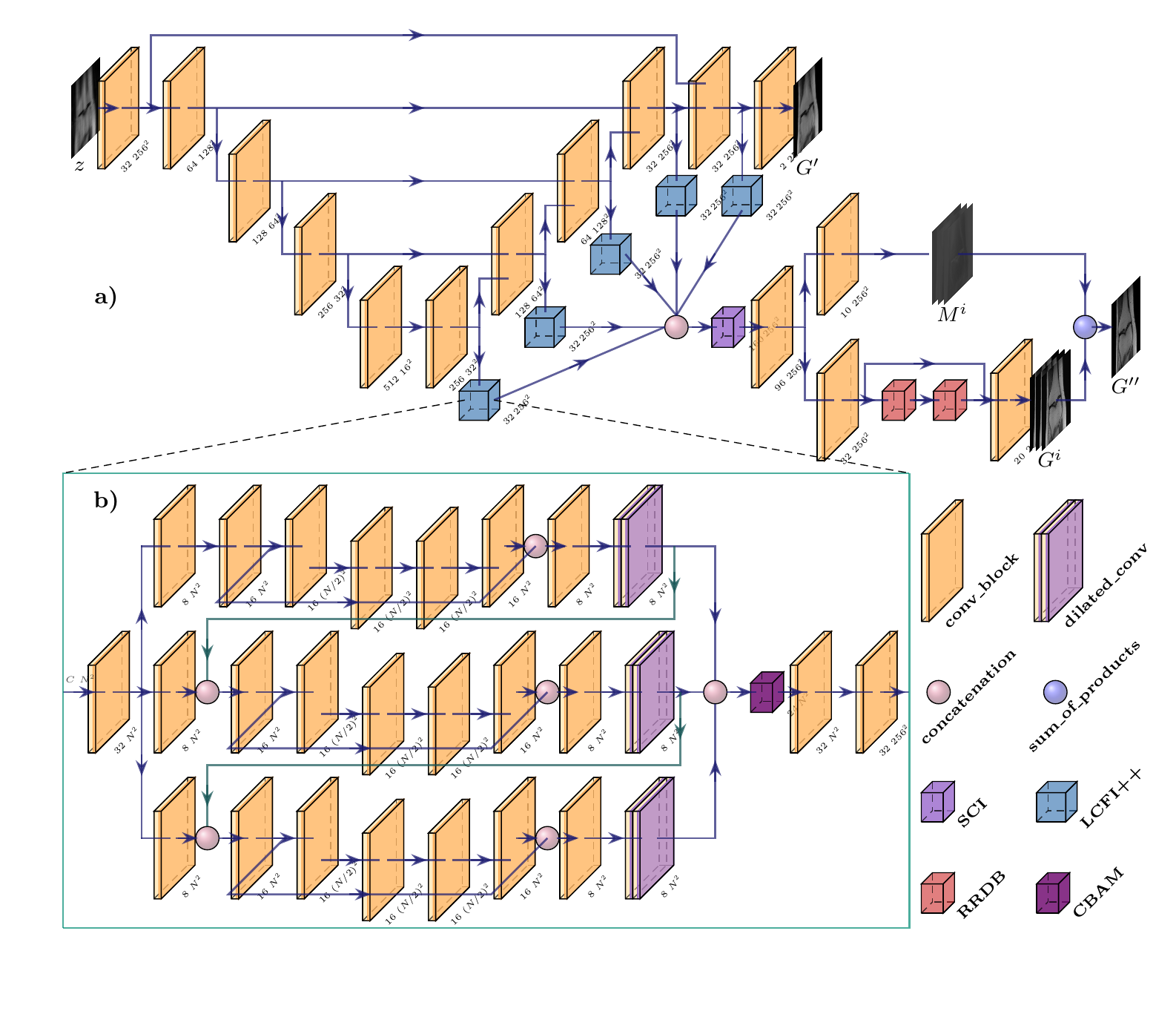}}
    \end{minipage}

    \caption{Overview of the model architecture, a) zero-filled is preliminarily recovered by the U-net and the features at different scales are used to give the final output by performing the temporal and spatial attentive selection, and b) the structure of LCFI++.}
    \label{fig:ASGAN_LCFI++}
\end{figure*}

\section{Method}
\subsection{Problem formulation}
Let $s$ be the fully-sampled MR image, i.e. sampled in $k$-space with Nyquist–Shannon rate. MRI reconstruction methods traditionally recover the measurement $y= \mathcal{H}\circ \mathcal{F}(s)$, where $\mathcal{H}$ and $\mathcal{F}$ respectively denote the under-sampling operation and Fourier transform, by the following equation,
\begin{equation}
    x = arg \min\limits_x \|y-\mathcal{H}\circ \mathcal{F}(x)\|^2 + \lambda R(x),
    \label{cs_mri}
\end{equation}
where $R(x)$ refers to a regularization penalty. However, existing optimization methods are computationally complex due to the iterative calculation of gradients, which involves manipulations of large matrices. We instead introduce a data-driven reconstruction framework which replaces the optimization process with a trained neural network, achieving high-quality reconstruction at a very low sampling rate.

\subsection{Model architecture}
We describe the architecture of proposed framework for MRI reconstruction in this section.

\subsubsection{LCFI++}
Motivated by the work in \cite{gdnet}, which introduces a novel large-field contextual feature integration (LCFI) module, to capture long-range dependencies, we propose a deep de-aliasing LCFI (LCFI++) block displayed in Figure \ref{fig:ASGAN_LCFI++}. We replace the spatially separable convolution with shallow U-nets in the parallel structure of LCFI. The outputs of dilated convolution layers are fused together by a convolutional block attention module (CBAM) \cite{cbam}. We discovered that our network equipped with LCFI++ behaves more stably than LCFI-integrated networks.

\subsubsection{Generator}
We propose to recover highly aliased observations in two stages, as illustrated in Figure \ref{fig:ASGAN_LCFI++}. Initially, we adopt a U-net as the backbone to produce the coarse reconstruction $G'$ from the aliased input. In order to circumvent an information bottleneck, we extract features from the predicted syntheses at different levels of the decoder by using the proposed LCFI++.

We exploit the multi-channel attention selection mechanism which performs a more sophisticated synthesis method as introduced in \cite{multi_channel_attention_selection_2020}. A self-channel interaction (SCI) block \cite{multi_channel_attention_selection_2020} is exploited to incorporate the channel-wise interdependencies among the outputs from LCFI++ modules. The resulting features are then used to construct multiple intermediate reconstructions $G^i$ and corresponding attention maps $M^i$. We adopt residual-in-residual dense blocks (RRDB) \cite{esrgan} in the generation of image maps to further enhance the reconstruction quality. $M^i$ and $G^i$ are later combined to produce the final output $G''$ as follows,
\begin{equation}
    G''= (M^1\otimes G^1)\oplus \cdots \oplus (M^N\otimes G^N),
    \label{attentive_selection}
\end{equation}
where $N$ is the number of attention pairs and $\otimes$ and $\oplus$ denote the element-wise multiplication and addition.

\subsubsection{Discriminator}
Conventional discriminators are trained to differentiate the synthetic date from real ones, which can fail to model local textures \cite{markovian_gan}. The Markovian discriminator is introduced in \cite{cyclegan} to encourage high-frequency components by convolutionally focusing on pixels in a fixed perceptual field, i.e. at the scale of patches. However, it is hard to select a suitable patch size in practice. Hence, we conform to the strategy in \cite{cyclegan} and utilise a multi-scale patch-based discriminator comprising 3 sub-networks with a shared structure to distinguish the patches at different scales. The discriminator is jointly trained with the generator in an adversarial learning diet.

\begin{figure*}[h]

    \begin{minipage}[b]{1.0\linewidth}
        \centering
        \centerline{\includegraphics[width=15.35cm]{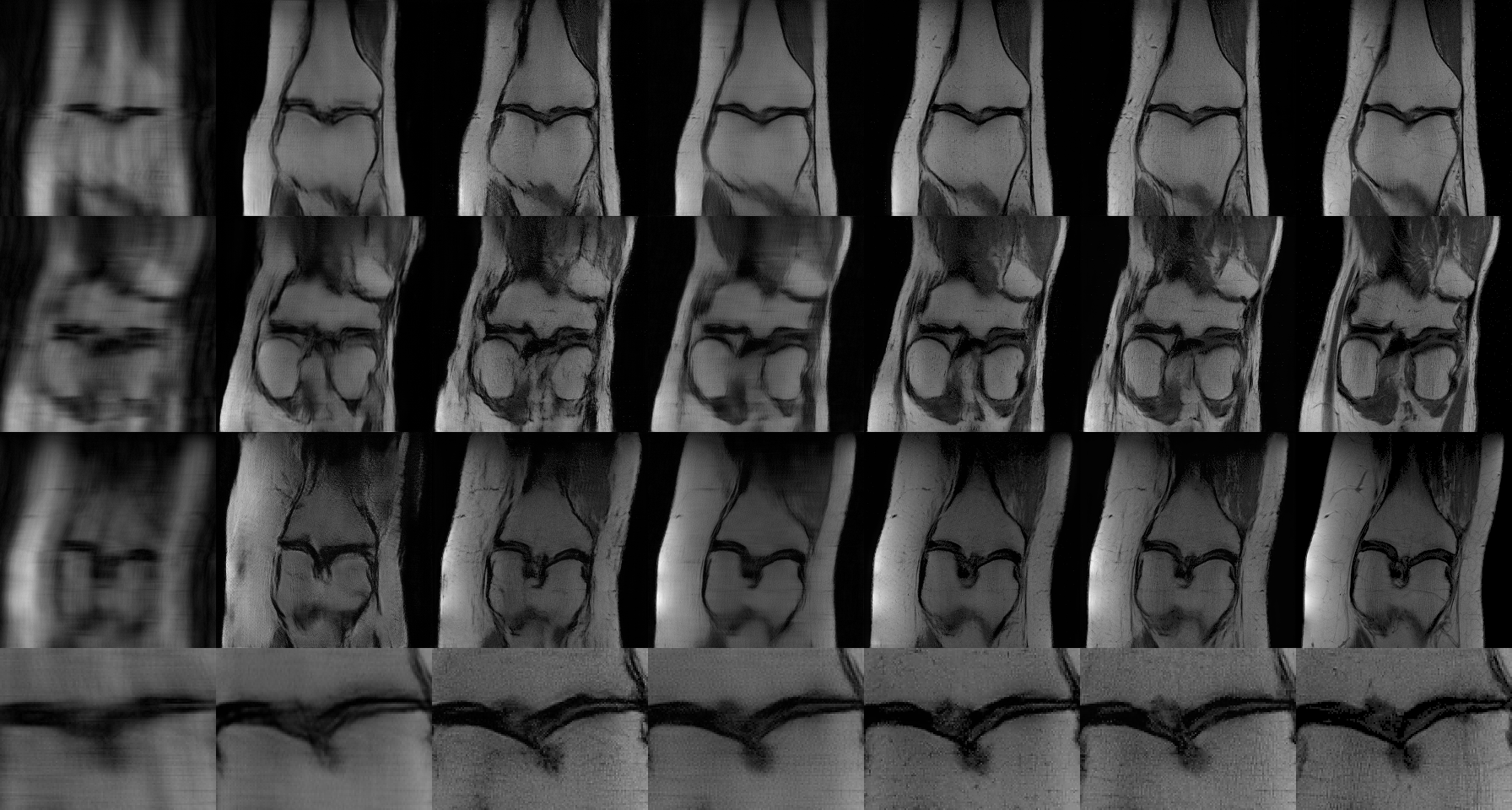}}
    \end{minipage}

    \caption{Comparison of 8$\times$ accelerated MRI reconstructions; first) zero-filled, second) CycleGAN-unsup \cite{cyclegan_pls_inverse}, third) LCFI++ (ours) with half of the dataset, fourth) MICCAN \cite{miccan}, fifth) CycleGAN-sup \cite{cyclegan_pls_inverse}, sixth) LCFI++ (ours), and last) ground truth.}
    \label{fig:experiment_0_1}
\end{figure*}

\subsection{Objective function}
We describe the loss functions used in the training phase in the following. We arithmetically combine the pixel-wise metric and structure-based index with the GAN-based objective. We encourage visually appealing results via a perceptual metric.

\subsubsection{Reconstruction loss}
We leverage the $L_1$-norm and multi-scale structural similarity index metric (MS-SSIM) \cite{ms_ssim} to enforce the generation to be favorable to the fully-sampled reference $s$. MS-SSIM is proved to be effective to preserve the contrast of high-frequency components, while $L_1$ can maintain luminance. The mixed loss function is then given as follows,
\begin{equation}
    \begin{split}
        L_{rec}=& \lambda_{rec}((1-\alpha)(L_1(G'',s) + \omega \sum L_1(M^i\otimes G^i, \\
        & M^i\otimes s)) +\alpha L^{SSIM}(G'',s) + \beta L_1(G',s)).
    \end{split}
    \label{loss_recons}
\end{equation}

\subsubsection{Adversarial loss}
Generative adversarial networks (GAN) \cite{vallina_gan} are proved to be effective to synthesize photo-realistic images, by leveraging a discriminator to differentiate real data from their synthetic counterparts and a generator to fool the discriminator. We use a least squares GAN (LSGAN) \cite{lsgan} to prevent the saturation during training. Empirically, LSGAN shows more stable reconstructions and faster convergences compared to Vallina GAN. The loss function is shown below,
\begin{align}
    L_{adv}^D &= E[\|D(s)-b\|_2^2] + E[\|D(G'')-a\|_2^2] \notag\\
    L_{adv}^G &= E[\|D(G'')-c\|_2^2],
    \label{loss_lsgan}
\end{align}
where $D$ denotes the discriminator, and the hyper-parameters $b$, $a$, and $c$ are selected to be $b=c=1$ and $a=0$.

\subsubsection{Data consistency}
Data consistency aims to enforce the outputs of the trainable model to agree with the observed measurements. We compute a consistency loss in $k$-space as it was also suggested in \cite{mri_reconstruction_with_cyclic_losses}. An alternative is to apply it in image domain as in \cite{cyclegan_pls_inverse}. We found that the two methods are not fundamentally different and in simulations they behave very similarly. The consistency loss is given as follows,
\begin{equation}
    L_{cyc}= \lambda_{cyc}\|y - \mathcal{H}\circ \mathcal{F}(G'')\|_1.
    \label{loss_latent}
\end{equation}

\subsubsection{Perceptual loss}
In addition to the aforesaid losses, we introduce the perceptual loss to improve the visual quality of the generated images. We utilize a pre-trained network, often a VGG pipeline \cite{style_transfer_cnn}, to map the images into the feature spaces which are more consistent with the human visual judgement. The perceptual difference is given by,
\begin{equation}
    \begin{split}
        L_{vgg} =& \lambda_{vgg} \sum \|f_{vgg}^i(G'')-f_{vgg}^i(s)\|_2^2 \\
        &+ \gamma\|f_{gram}^i(G'')-f_{gram}^i(s)\|_2^2,
    \end{split}
    \label{loss_vgg}
\end{equation}
where $f_{vgg}^i$ denotes the pre-activations of the $i$-th layer in VGG and $f_{gram}^i$ represents their Gram matrix.

\section{Simulations}
We extract 2800 images from the NYU knee MRI database \cite{fastmri} for training and 164 samples for test. All images are resized to $256\times 256$. We use two channels to represent the real and imaginary parts of complex-valued images. We adopt a fixed random sampling mask at a rate of 12.5$\%$. We apply the inverse Fourier transform to $y$ to produce the zero-filled $z$, which is the input to the generator. Our code will be provided at https://github.com/JingshuaiLiu/ASGAN.
\begin{table}[h]
    \caption{Quantitative Evaluation}
    \label{table:evaluation}
    \begin{tabularx}{1.0\linewidth}{@{}l*{10}{C}c@{}}
        \toprule
        method     & PSNR$\uparrow$  & SSIM$\uparrow$ & FID$\downarrow$ & KID$\downarrow$  \\
        \midrule
        proposed   & 25.45      & 0.638          & 104.34   & 0.036   \\ 
        CycleGAN-sup & 25.62        & 0.655        & 143.36    & 0.093   \\ 
        MICCAN       & 26.61       & 0.642          & 180.66  & 0.146  \\ 
        CycleGAN-unsup        & 22.03       & 0.589          & 209.31   & 0.184  \\ 
        \bottomrule
    \end{tabularx}
\end{table}

\subsection{Qualitative analysis}
We compare the results of the proposed framework with other two state-of-the-art deep learning-based methods. A deep neural network with channel-wise attention modules (MICCAN) for MRI reconstruction is introduced in \cite{miccan}. We implement the method proposed in \cite{cyclegan_pls_inverse}, denoted by CycleGAN-unsup, which trains a CycleGAN \cite{cyclegan} to recover the measurements in an "unsupervised" manner. We also present the outputs of the CycleGAN-based framework trained under supervision (CycleGAN-sup), i.e. using the whole fully-sampled dataset, which is presented in \cite{cyclegan_pls_inverse} as a reference method.

To further testify the performance of the proposed model, we train the pipeline only with half the training data and compare it with CycleGAN-unsup which uses the same number of high-quality samples. As CycleGAN-unsup has access to the other half of database in low-resolution, a comparison with the proposed method using only half of the dataset is fair or in the favour of CycleGAN. We display all comparison results in Figure \ref{fig:experiment_0_1}. We can observe that the proposed model produces sharper and more detailed reconstructions than CycleGAN-unsup and MICCAN, and the generated images are more natural and relevant to full reconstructions than CycleGAN-sup. In the zoomed view, i.e. bottom row, we can observe that the proposed framework recovers finer textures, which confirms that our method can achieve superior fine-grained reconstruction with more complicated local details in those examples, while preserving the global and structural information.

\subsection{Quantitative analysis}
We draw a quantitative analysis using PSNR and SSIM as evaluation metrics. We adopt the Fr\'{e}chet inception distance (FID) and kernel inception distance (KID) \cite{kid} to measure the visual quality of reconstructions. Table \ref{table:evaluation} shows their average scores over all test samples. Overall, our model is competitive with CycleGAN-sup and MICCAN in terms of PSNR and SSIM, in which higher figures are better, and produces superior results than all the other methods with remarkably better FID and KID, in which lower figures are better.

\section{Conclusions and Discussion}
In this paper, we introduce a novel deep learning-based model to achieve fine-grained MRI reconstruction. We leverage a long-range contextual feature integration module with attention selection mechanism in a GAN-based framework to recover the observed measurements. We demonstrate that the proposed framework outperforms other deep neural network-based methods in terms of image quality. The qualitative and quantitative analyses prove its usefulness in providing accurate and fine-grained reconstruction. The future works include incorporating the proposed method with model-based algorithms to provide the convergence guarantee and extending the model to dynamic and parallel imaging.

\bibliographystyle{IEEEbib_2021}
\bibliography{bibfile}

\end{document}